\documentstyle[epsfig,prl,aps]{revtex}

\begin{document}
\draft \twocolumn[\hsize\textwidth\columnwidth\hsize\csname
@twocolumnfalse\endcsname
\title{$\Lambda_{\nu}^{\mu}$ geometries from the point of view of different
observers}
\author{Irina Dymnikova
}
\address{Department of Mathematics and Computer Science,
         University of Warmia and Mazury,\\
Zolnierska 14, 10-561 Olsztyn, Poland; e-mail:
irina@matman.uwm.edu.pl}

\maketitle

\begin{abstract}
$\Lambda^{\mu}_{\nu}$-geometry is a geometry with a variable
cosmological term described by a second-rank symmetric tensor
$\Lambda^{\mu}_{\nu}$ whose asymptotics are Einstein cosmological
term $\Lambda \delta ^{\mu}_{\nu}$ at the origin and $\lambda
\delta ^{\mu}_{\nu}$ at infinity (with $\lambda < \Lambda$).
 It corresponds to extension of the algebraic
structure of the Einstein cosmological term $\Lambda \delta
^{\mu}_{\nu}$ in such a way that a scalar $\Lambda$ describing
vacuum energy density as $\rho_{vac}=8\pi G \Lambda$ (with
$\rho_{vac}$=const by virtue of the Bianchi identities), becomes
explicite related to the appropriate component, $\Lambda^0_0$, of
an appropriate stress-energy tensor, $T^{\mu}_{\nu}=8\pi
G\Lambda^{\mu}_{\nu}$ whose vacuum properties follow from its
symmetry, $T_0^0=T_1^1$, and whose variability follows from the
contracted Bianchi identities. In the spherically symmetric case
existence of such geometries in frame of GR follows from imposing
on Einstein equations requirements of finiteness of the ADM mass
$m$, and of regularity of density and pressures. Dependently on
parameters $m$ and $q=\sqrt{\Lambda /\lambda}$,
$\Lambda^{\mu}_{\nu}$ geometry describes five types of
configurations. We summarize here the results which tell us how
these configurations look from the point of view of different
observers: a static observer, a Lemaitre co-moving observer, and a
Kantowski-Sachs observer.

\vskip0.1in

 $^{*}$ {\bf Talk at the Fourth
International Conference on Physics Beyond the Standard Model
"Beyond the Desert'03", Castle Ringberg, Germany, June 2003; to
appear in "Physics Beyond the Standard Model, BEYOND'03", Ed. H.V.
Klapdor-Kleingrothaus}

\end{abstract}

\vskip0.2in

 ]

\section{Introduction}

The motivation for introducing a geometry able to model
Lambda-variability is evident: With great probability we live in
$\lambda$-dominated universe with $\lambda$ contributing via
$\rho_{vac}=8\pi G \lambda$ to the total energy density  with
$\rho_{vac}\simeq {0.73 \rho_{total}}$ \cite{Bahcall}. On the
other hand, the big value of Lambda is responsible for an earliest
inflationary stage solving various puzzles of the standard
cosmology \cite{olive}.

The Einstein cosmological term corresponds to a maximally
symmetric vacuum stress-energy tensor \cite{gliner,dewitt}
$$
\Lambda \delta^{\mu}_{\nu}=8\pi G \rho_{vac} \delta^{\mu}_{\nu}
                                                 \eqno(1)
$$
A variable cosmological term is introduced as an extension of
$\Lambda \delta^{\mu}_{\nu}$ to a second-rank symmetric tensor
$\Lambda^{\mu}_{\nu}$ connecting smoothly two asymptotic maximally
symmetric states (1) with different values of cosmological
constant \cite{me2000} (for review \cite{merev,me2003})
$$
\Lambda \delta^{\mu}_{\nu} \leftarrow \Lambda^{\mu}_{\nu}
\rightarrow \lambda \delta^{\mu}_{\nu}
                                          \eqno(2a)
$$
corresponding to certain vacuum tensor
$$
\rho_{vac}^{in} \delta^{\mu}_{\nu} \leftarrow T^{\mu}_{\nu}
\rightarrow \rho_{vac}^{out} \delta^{\mu}_{\nu}
                                          \eqno(2b)
$$
 Kind of justification for
identification of eq.(2) comes from nice discussion in
Ref.\cite{overduin} about where to put the cosmological term in
the Einstein equation
$$
G^{\mu}_{\nu} + \Lambda \delta^{\mu}_{\nu} =0
                                                                \eqno(3)
$$
According to \cite{overduin} putting it on the right-hand side of
the Einstein equation as $T^{\mu}_{\nu}=8\pi G \rho_{vac}
\delta^{\mu}_{\nu}$ and treating as a dynamical part of a matter
content, originates from dialectic materialism of the Soviet
physics school where it came from \cite{gliner,zeld}, although the
first calculation relating $\Lambda \delta^{\mu}_{\nu}$ to a QFT
vacuum was done by De Witt \cite{dewitt}. On contrary keeping
$\Lambda \delta^{\mu}_{\nu}$ on the left-hand side and treating it
as geometrical entity with $\Lambda$ as a fundamental constant of
nature, is preferred by idealistic approach.

The first approach - shifting $\Lambda \delta^{\mu}_{\nu}$ to the
right side as some $T^{\mu}_{\nu}$, allows energy exchange of de
Sitter vacuum (1) with another matter.  Even a simplest
nonsingular FRW cosmological model with the initial de Sitter
state and conversion of a vacuum energy into radiation, gives
(generically, for any source of $\Lambda$) needed entropy and
accelerated exponential expansion with e-folding number sufficient
to explain homogeneity, isotropy, present size and density of the
Universe \cite{us75} (for review \cite{olive}).

Such an approach is most popular till now for making
$\Lambda$-variability due to exchange with other matter in various
models  \cite{perf,overduin}, and for solving problem of
cosmological constant in field theories by cancelling pre-existing
vacuum energy (cosmological constant) with developed by an
appropriate field big expectation value of a vacuum density
\cite{cancel} (for review \cite{canrev}).

The aim of the reported research was to reveal how one could make
cosmological term variable in itself.

The approach in introducing (2) was in a sense opposite to that
classified as materialistic. We first found in the right-hand side
of the Einstein equation
$$
G^{\mu}_{\nu}=-8\pi G T^{\mu}_{\nu}
                                                \eqno(4)
$$
the class of source terms with the proper symmetry
\cite{me92}
$$
T^t_t=T^r_r; ~~~ ~~~~T^{\theta}_{\theta}=T^{\phi}_{\phi}
                                                              \eqno(5)
$$
 and asymptotic behavior \cite{us97}
$$
\Lambda \delta^{\mu}_{\nu} \leftarrow 8\pi G T^{\mu}_{\nu}
\rightarrow \lambda \delta^{\mu}_{\nu}
                                          \eqno(6)
$$
shift it then to the left-hand side of equation (4) as
\cite{me2000}
$$
\Lambda^{\mu}_{\nu} = 8\pi G T^{\mu}_{\nu}
                                            \eqno(7)
$$
treating as evolving and clustering geometrical entity.

The cosmological term (7) is invariant under radial boosts and is
identified as corresponding to a spherically symmetric anisotropic
vacuum \cite{me92}, with the algebraic structure  [(II)(II)] in
the Petrov classification.

The existence of the class of regular spherical metrics generated
by stress-energy tensors of structure (5) follows from
requirements of regularity of density $=\rho$ and finiteness of
the ADM mass $m$, and imposing  dominant energy condition on a
stress-energy tensor \cite{me2002,me2003}, either weak energy
condition and regularity of pressures \cite{stab}. These two
possibilities correspond to geometries with variable curvature
scalar $R$, positive in the first case \cite{stab}.

 At present there is known a lot of matter sources contributing to GR
equations with a stress-energy tensor of structure (1). All of
them give the same geometry - de Sitter geometry governed by
$\Lambda \delta ^{\mu}_{\nu}$ - whose generic properties are used
then in relevant physical models \cite{olive}.

 In similar way we study geometries generated by (7)
 whose mathematical properties are
 generic.
 The advantage is using of possibilities given by mathematical
models followed from GR equations, to obtain a Lambda-dominated
stage at which a matter comes into play.

In this talk we show how the vacuum configurations represented by
$\Lambda^{\mu}_{\nu}$ geometry, look from the point of view of
different observers: a static observer, a co-moving Lemaitre
observer and a Kantowski-Sachs observer.

\section{$\Lambda^{\mu}_{\nu}$ geometry}

The Einstein cosmological term (1) is identified as a vacuum
stress-energy tensor due to its maximally symmetric form.  It is
invariant under any coordinate transformation which makes
impossible to distinguish a preferred co-moving reference frame
\cite{gliner}. As a result an observer moving through a medium
with a stress-energy tensor (1) cannot in principle measure his
velocity with respect to it, which allows one  to classify it as a
vacuum in accordance with the relativity principle \cite{gliner}.

Introducing in similar way a vacuum with variable density, one
cannot keep the full invariance which leads, by Bianchi
identities, to $\rho_{vac}=$ const. The invariance can be kept for
an observer moving along a certain direction in  space
distinguished by symmetry of a source term (5). In
$\Lambda^{\mu}_{\nu}$ extension of $\Lambda \delta^{\mu}_{\nu}$ a
constant scalar $\Lambda$ associated by (1) with a vacuum density,
constant by virtue of Bianchi identities, becomes a tensor
component $\Lambda^t_t= 8\pi G T^t_t$ associated explicitly with a
density component of a perfect fluid stress-energy tensor, whose
vacuum properties follow from its symmetry and whose variability
follows just from the Bianchi identities \cite{me2000}.

For the stress-energy tensor of the algebraic structure (5) the
generalized Birkhoff theorem \cite{birk} guarantees the existence
of a coordinate frame where the metric has the static form

$$
ds^2=g(r) dt^2- \frac{dr^2}{g(r)} - r^2 d\Omega^2
                                                    \eqno(8)
$$
where $d\Omega^2$ is the line element on the unit 2-sphere. Here
we concentrate on the spherically symmetric case although
generalization is straightforward to the cases of 2-dimensional
planes and Lobachevsky planes \cite{us03}.

For the case of $\Lambda^{\mu}_{\nu}$ with asymptotic behavior
(2), the metric function is given by \cite{us97}
$$
g(r)=1-\frac{R_g(r)}{r} - \frac{\lambda}{3} r^2
                                                   \eqno(9)
$$
with
$$
R_g(r)=2 G M(r);~~M(r)=4\pi \int_0^r{\rho(r) r^2 dr}
                                              \eqno(10)
$$
The asymptotic value of lambda at infinity is included into
stress-energy tensor in such a way that
$$
T^t_t=\rho(r) + (8 \pi G)^{-1}\lambda
                                                \eqno(11)
$$
so that the ADM mass $m$ is defined in the standard way
$$
m=4\pi \int_0^{\infty}{\rho(r) r^2 dr}
                                             \eqno(12)
$$

From the conserved Bianchi identities it follows the conservation
equation $\Lambda^{\mu}_{\nu;\mu}=0$ which gives the equation of
state connecting components of a cosmological term
$\Lambda^t_t=8\pi G \rho(r)$,
 $\Lambda^r_r=-8\pi G p_r$, $\Lambda^{\theta}_{\theta}=\Lambda^{\phi}_{\phi}
 =-8\pi G p_{\perp}$ \cite{me2000}
$$
p_r=-\rho; ~~ ~p_{\perp}=p_r+\frac{r}{2}\frac{dp_r}{dr}
                                                          \eqno(13)
$$

In the case of two scales for vacuum density, geometry has  not
more than three horizons \cite{us03}, and describes five types of
configurations as shown in Fig.1 \cite{us97}.

\begin{figure}
\vspace{-8.0mm}
\begin{center}
\epsfig{file=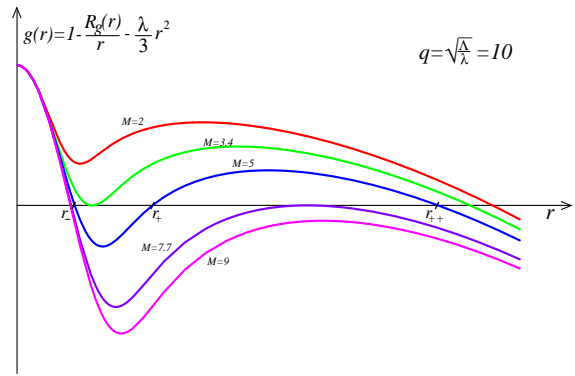,width=8.0cm,height=5.5cm}
\end{center}
\caption{ The metric $g(r)$ for $\Lambda^{\mu}_{\nu}$
configurations.
The mass $m$ is normalized to $(3/G^2\Lambda)^{1/2}$.
The parameter $q=\sqrt{\Lambda/\lambda}$. }
\label{fig.1}
\end{figure}

The case of three horizons, an internal horizon $r_{-}$, a black
(white) hole horizon $r_{+}$, and a cosmological horizon $r_{++}$,
corresponds to the nonsingular modification of Kottler-Treftz
solution \cite{kot} referred to in the literature as
Schwarzschild-de Sitter geometry \cite{gh}. In the regular version
it exists only within a certain range of mass parameter, $m_{min}
\leq m \leq m_{max}$, with limits depending on the value of the
parameter $q=\sqrt{\Lambda/\lambda}$ \cite{us97}.

In $\Lambda^{\mu}_{\nu}$ geometries singularities $r=0$  are
replaced with a regular R-regions asymptotically de Sitter with
the value of $\Lambda$ as $r\rightarrow 0$ corresponding to a
scale of symmetry restoration \cite{me97,me2002}. In these regions
$g(r)>0$. The regions where $g(r) < 0$ are called T-regions.

R- and T- regions are specified by the invariant quantity
$\Delta=g^{\mu\nu}r_{,\mu}r_{,\nu}$ (see, e.g., \cite{igor}). In
R-regions $\Delta < 0$, the surfaces $r$=const are time-like,
static observers can exist, move and send signals in both
directions. In T-regions $\Delta > 0$, the surfaces $r$=const are
space-like, no static observer can exist in principle, both
signals from this surface propagate in the same direction.
T-regions are regions of one-way traffic. In Fig. 1 they are
located between horizons $r_{-},~r_{+}$ and between $r_{++}$ and
infinity. At horizons $\Delta=0$. For the metric in the Kruskal
form $\Delta = (1/2) g(r)^{-1} r_{,u}r_{,v}$, and the conditions
$r_{,u}
> 0$ and $r_{,u} < 0$ are invariant: $r_{,u} < 0$
for contracting $T_{-}$ region; $r_{,u} > 0$ for an expanding
$T_{+}$ region.

\vskip0.1in

{\bf Static observers}

Static observers exist only in R-regions. In all types of
configurations horizons exist, at least one related to replacing a
singularity with a de Sitter core. In most general case of three
horizons (see Fig.1), a static observer between $r_{+}$ and
$r_{++}$ observes a vacuum nonsingular cosmological black (white)
hole, and his horizons are black hole horizon $r_{+}$ and
cosmological horizon $r_{++}$ in his future (past). A static
observer between $r=0$ and $r=r_{-}$ sees internal horizon $r_{-}$
as his cosmological (future or past) horizon. It is seen from
Fig.2 which presents the global structure of $\Lambda_{\mu\nu}$
geometry with three horizons \cite{us97}.
\begin{figure}
\vspace{-8.0mm}
\begin{center}
\epsfig{file=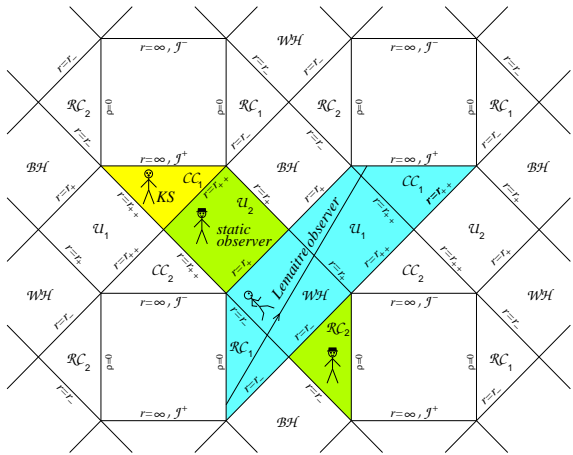,width=8.0cm,height=6.5cm}
\end{center}
\caption{ The global structure of space-time
 with 3 horizons.}
\label{fig.2}
\end{figure}

Static observers see Hawking radiation from black hole horizons,
and Gibbons-Hawking radiation \cite{gh} from  cosmological
horizons, with the temperature \cite{me96}
$$
k T=\frac{\hbar c}{4\pi} \biggl[\frac{R_g(r_h)}{r_h^2}-
\frac{R_g^{\prime}(r_h)}{r_h}\biggr]; ~~~ r_h=r_{-}, r_{+}, r_{++}
$$

\vskip0.1in

{\bf Kantowski-Sachs observers}

In T-regions the coordinates $r$ and $t$ interchange their roles,
$r$ becomes time-like and $t$ space-like. The metric (8) can be
re-written, introducing time coordinate $u$, space-like coordinate
$x$ and the metric function is $a^2(u)=-g(u)$, in the form
\cite{us03}
$$
ds^2 = \frac{1}{a^2(u)} du^2 - a^2(u) dx^2 - u^2 d\Omega ^2
                                                                 \eqno(14)
$$
It describes an anisotropic cosmological model with two
time-dependent scale factors $a(u)$ and $b(u)=u$, and the lapse
function $1/a^2(u)$. A spatial section has topology of a
3-dimensional cylinder with different time-dependent scale factors
in the radial and longitudinal directions \cite{us03}.

Models described by metric (14) belong to  Kantowski-Sachs  type
homogeneous anisotropic cosmological models with two
time-dependent scale factors \cite{kant}. Kantowski-Sachs models
represent a special class of T-models \cite{Ruban} which are in
general inhomogeneous.

 Essentially new feature of these
models in our case is the existence of regular R-region near
$r=0$, so metrics (14) are identified as Kantowski-Sachs models
with the regular R-regions \cite{us03}.\footnote{In the case of
pseudospherical symmetry 2-space is Lobachevsky plane
        models are identified as hyperbolic Kantowski-Sachs models \cite{us03}.
        In the case of the planar symmetry of 2-space, models belong
        to Bianchi type I \cite{us03}.}

These models can describe both regular cosmologies in
$T_{+}$-regions and regular collapse in $T_{-}$-regions.

For Kantowski-Sachs observers the cosmological evolution starts
from horizons $u=r_h$ which are highly anisotropic purely
coordinate singularities (the 4-geometry is perfectly globally
regular), where coordinate surfaces, spheres with the same finite
scale factor $u(r_h)$ stick to one another. As a result
cosmological evolution for Kantowwski-Sachs observers starts with
a {\bf null bang from a horizon}, a null surface ($u=r_h$) with
vanishing volume of spatial section squeezed along $x (a(r)=0)$;
this happens at finite cosmological time $\tau(u)=\int{du/a(u)}$
for the case of a simple horizon, and in the infinitely remote
past for higher-order horizons \cite{us03}.

In our case Kantowski-Sachs models have regular R-regions, and
Kantowski-Sachs observers can receive information from their
remote past (even in the case of high-order horizons) brought by
particles and photons crossing a horizon in their finite proper
time \cite{us03}.

\vskip0.1in

{\bf Lemaitre observers}

As usual in space-times with horizons, a static observer can see
only small part of the manifold, R-region in which he is actually
resided. The same concerns Kantowski-Sachs observer who  exists
only in T-regions. Both have problems with horizons which are
singular surfaces for them. Removing coordinate singularities in
the way similar to applied by Lemaitre for Schwarzschild geometry,
one introduces coordinates of an observer to whom more extended
parts of a manifold are available.

Connecting coordinates with the particles moving on radial
geodesics marked by the constant of motion $E^2$, we make
transformation $r,t ~~\rightarrow ~~R,\tau$ where $R,\tau$ are
coordinates of  co-moving observers in which the metric (8) takes
the Lemaitre form \cite{us03}
$$
ds^2 =d\tau^2 -\frac{(E^2-g(r(R,\tau))}{E^2}dR^2 -
r^2(R,\tau)d\Omega^2
                                                             \eqno(15)
$$
  The models described by (15), belong to the Lemaitre class of cosmological models
with anisotropic perfect fluid, since the principal pressures are
essentially different for $\Lambda^{\mu}_{\nu}$ geometry
satisfying the equation of state (13). In coordinates connected
with in-falling particles, the metric (15) describes a regular
gravitational collapse.

For all $\Lambda^{\mu}_{\nu}$ configurations, Lemaitre class
cosmologies describe evolution starting with {\bf a nonsingular
non-simultaneous de Sitter bang} \cite{us2001,us03}.

It can be easily seen for the case of de Sitter-Schwarzschild
geometry \cite{me96} which is the particular case of
$\Lambda^{\mu}_{\nu}$ geometry with $\lambda=0$.
 The  global structure of de Sitter-Schwarzschild space-time
is shown in Fig.3 where it is compared with the case of
Schwarzschild geometry.
\begin{figure}
\vspace{-8.0mm}
\begin{center}
\epsfig{file=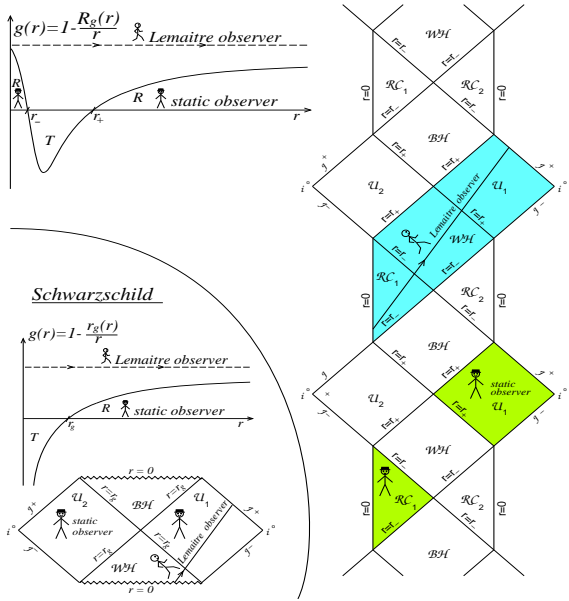,width=8.0cm,height=9.5cm}
\end{center}
\caption{ De Sitter-Schwarzschild space-time. }
\label{fig.3}
\end{figure}

Lemaitre coordinates $R, \tau$ map the segment $\cal{RC}$ (regular
core) , $\cal{WH}$ (white hole), $\cal{U}$ (universe) available to
Lemaitre observers. A regular core $\cal{RC}$ models an initial
state for an expanding universe in all $\Lambda^{\mu}_{\nu}$
configurations. Evolution starts from the surface $r(R, \tau)=0$
which is the bang surface. In Schwarzschild case this is the
surface of big bang singularity \cite{silk}. For example, in the
case of the Tolman-Bondi dust model, it is described by
$r(R,\tau)=(9M(R)/2)^{1/3}(\tau-\tau_0(R))^{2/3}$ with the
bang-time function $\tau_0(R)$ \cite{tb}.

In the case of Schwarzschild white hole the bang surface $c\tau +
R =0$ (see Fig.4) is space-like surface (see Fig.3). In de
Sitter-Schwarzschild geometry the bang surface ($c\tau + R =
-\infty$ in Fig.4) is the time-like regular surface ($r=0$ in
Fig.3).
 \begin{figure}
\vspace{-8.0mm}
\begin{center}
\epsfig{file=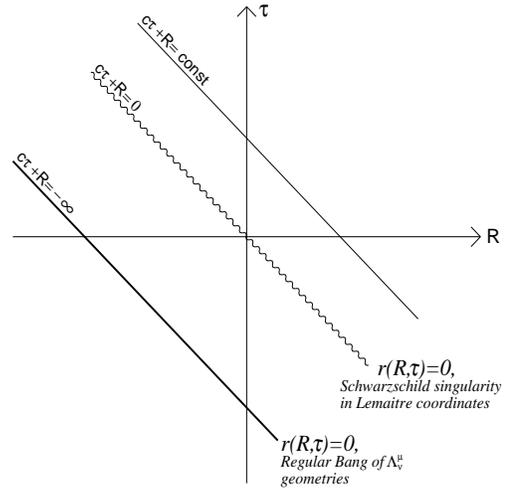,width=7.0cm,height=7.0cm}
\end{center}
\caption{ The surfaces $r$=const in Lemaitre coordinates.}
\label{fig.4}
\end{figure}
Since a bang surface is timelike, different points  start at
different moments of the synchronous time $\tau$.

 Near the bang surface the metric (15) takes
the FRW form
$$ ds^2 = d\tau^2 -a^2(\tau)(d\chi^2 + sin^2 \chi d\Omega^2)
                                                          \eqno(16)
$$
with the de Sitter scale factor $a(\tau)\sim{\exp{(H\tau)}}$ for a
spatially flat model, $a(\tau) \sim {\cosh (H\tau)}$ for a closed
model, and $a(\tau) \sim {\sinh(H\tau)}$ for an open model; $H$ is
the Hubble parameter $H=\sqrt{\Lambda/3}$ \cite{us2001}.

An inflationary stage is followed
 by a Kasner-type anisotropic stage, with contraction
     in the radial direction and expansion in the tangential direction, at
     which most of a universe mass is produced \cite{us2001}.

 The metric (15)
     at the Kasner-type stage takes the form \cite{us2001,us03}
$$
     ds^2 = d\tau^2- (\tau {+} R)^{-2/3} F(R) dR^2 -
B(\tau {+} R)^{4/3} d\Omega_1^2
                                                            \eqno(17)
$$
     where $F(R)$ is a smooth regular function and $B$ is a constant
     related to the model parameters. At this stage acceleration is changes
drastically (see Fig.5 \cite{us2001}).
\begin{figure}
\vspace{-8.0mm}
\begin{center}
\epsfig{file=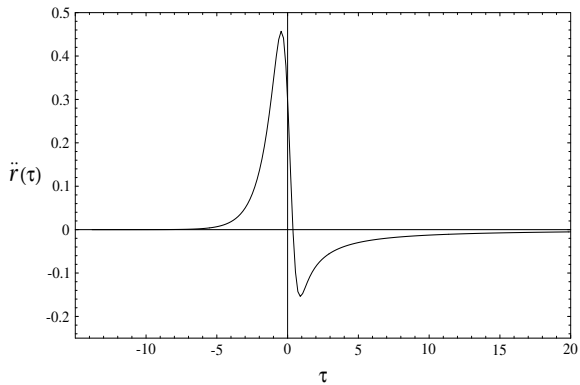,width=8.0cm,height=5.5cm}
\end{center}
\caption{Acceleration of the scale factor $r(\tau)$
\protect\cite{us2001}.}
\label{fig.5}
\end{figure}
At late times all $\Lambda^{\mu}_{\nu}$ dominated models become
isotropic and approach de Sitter asymptotic described by (16) with
the Hubble parameter $H=\sqrt{\lambda/3}$ \cite{us03}.

In the case $E^2=1$ models  asymptotically approach flat FRW
models in both past $\Lambda$ and future $\lambda$ limits; a de
Sitter bang occurs in infinitely remote past, $R+c\tau=-\infty$.
In the case $E^2
> 1$, when the initial velocity on the geodesics of the reference
frame is nonzero, models asymptotically approach open FRW models;
a bang occurs at finite $R+\tau$ \cite{us03}. In case $E^2 < 1$
(asymptotically approaching closed FRW model) a bang starts from a
finite value $r_{in}$ given by $E^2-g(r_{in})=0$ at finite
$R+\tau$.

\section{Possibilities given by $\Lambda_{\mu\nu}$ geometry}

\vskip0.1in

 {\bf One horizon configurations}

\vskip0.1in

 There are two types of one-horizon configurations
shown in Fig.6. In both cases the global structure of space-time
is the same as for de Sitter geometry. Essential difference is
that cosmological density $\Lambda^t_t$ in $\Lambda_{\mu\nu}$
geometry evolves smoothly from $\Lambda$ to $\lambda$. This is
evolution in $r$ for a static observers in R-region, in proper
time $\tau$ for Lemaitre and Kantowski-Sachs observers.

For Lemaitre observers evolution starts from non-simultaneous de
Sitter bang $r(R,\tau)=0$. For Kantowski-Sachs observers evolution
starts with a null bang from a single horizon, in their finite
proper time.

\begin{figure}
\vspace{-8.0mm}
\begin{center}
\epsfig{file=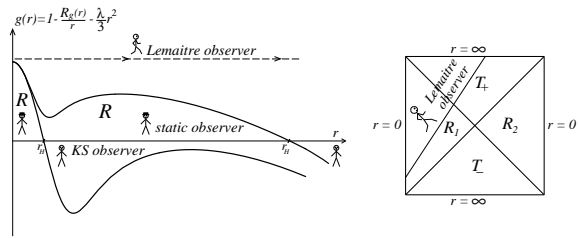,width=8.0cm,height=3.5cm}
\end{center}
\caption{$\Lambda^{\mu}_{\nu}$ geometry with one horizon. }
\label{fig.6}
\end{figure}

In the first case, the upper curve in Fig.6, the horizon is
related to the small asymptotic value, $\lambda$, $r_{-}\approx
{\sqrt{\lambda/3}}$. Observers in R-region see Hawking radiation
with the Gibbons-Hawking temperature $T \sim {\hbar c
\sqrt{\lambda/3}}$. The essential dynamical changes occur in the
R-region, experienced by static observers, or by comoving Lemaitre
observers before crossing horizon.

In second case, the lower curve in Fig.6, the horizon is
approximately related to big asymptotic value, $\Lambda$,
$r_{++}\approx{\sqrt{\Lambda/3}}$. Observers see Hawking radiation
with the Gibbons-Hawking temperature  $T \sim {\hbar c
\sqrt{\Lambda/3}}$. One can say that they are in much hotter
environment.  The essential dynamical changes occur in T-region as
dynamical possibilities for Kantowski-Sachs observers.

\vskip0.1in

{\bf Configuration with three horizons}

\vskip0.1in

This configuration is shown in Fig.7 \cite{us97}.
\begin{figure}
\vspace{-8.0mm}
\begin{center}
\epsfig{file=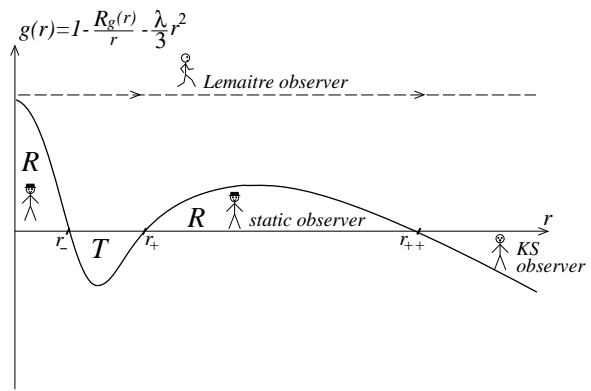,width=8.0cm,height=5.5cm}
\end{center}
\caption{ $\Lambda^{\mu}_{\nu}$ geometry with three horizons. }
\label{fig.7}
\end{figure}
The global structure of spacetime is shown in Fig.2.

An observer in R-region between $r_{+}$ and $r_{++}$ observes
Hawking radiation from both black hole and cosmological horizons,
whose temperature in certain range of parameters  can be estimated
as $kT \sim {\hbar c^3 /8\pi G m}$ for black hole horizon and by
$kT\sim {\hbar c \sqrt{\lambda/3}}$ for a cosmological horizon. An
observer in R-region between $r=0$ and $r=r_{-}$ observes
Gibbons-Hawking radiation with temperature estimated roughly by
$kT \sim {\hbar c \sqrt{\Lambda/3}}$.

The segments $\cal{RC}, \cal{WH}, \cal{U}, \cal{CC}$, are
available to Lemaitre observers. They can cross horizons without
big problems (tidal forces are big only when $m$ is small).
Coordinate frame related to radially infalling observers describes
nonsingular contraction, while frames related to outgoing
geodesics describe cosmological models of Lemaitre class with a
non-simultaneous de Sitter bang.

T-regions of a black (white) hole between $r_{-}$ and $r_{++}$ and
cosmological regions between $r_{++}$ and infinity are regions of
one-way traffic. The cosmological regions are the place of
Kantowski-Sachs observers. From their point of view evolution
starts from a horizon $r_{++}$ with a null bang in their remote
past but in finite proper time.

Even in such a complicated space-time KS observers can receive
information about pre-bang history brought by particles and
photons from the first R-region, first  $T_{+}$-region which in
this case is a white hole, and next R-region (universe of a static
observer).

 \vskip0.1in

{\bf Double horizon $r_{+}=r_{++}$}

\vskip0.1in

This configuration, the first extreme state of a nonsingular
cosmological black hole  (nonsingular modification of the Nariai
geometry \cite{nar}), is shown in Fig.8.
\begin{figure}
\vspace{-8.0mm}
\begin{center}
\epsfig{file=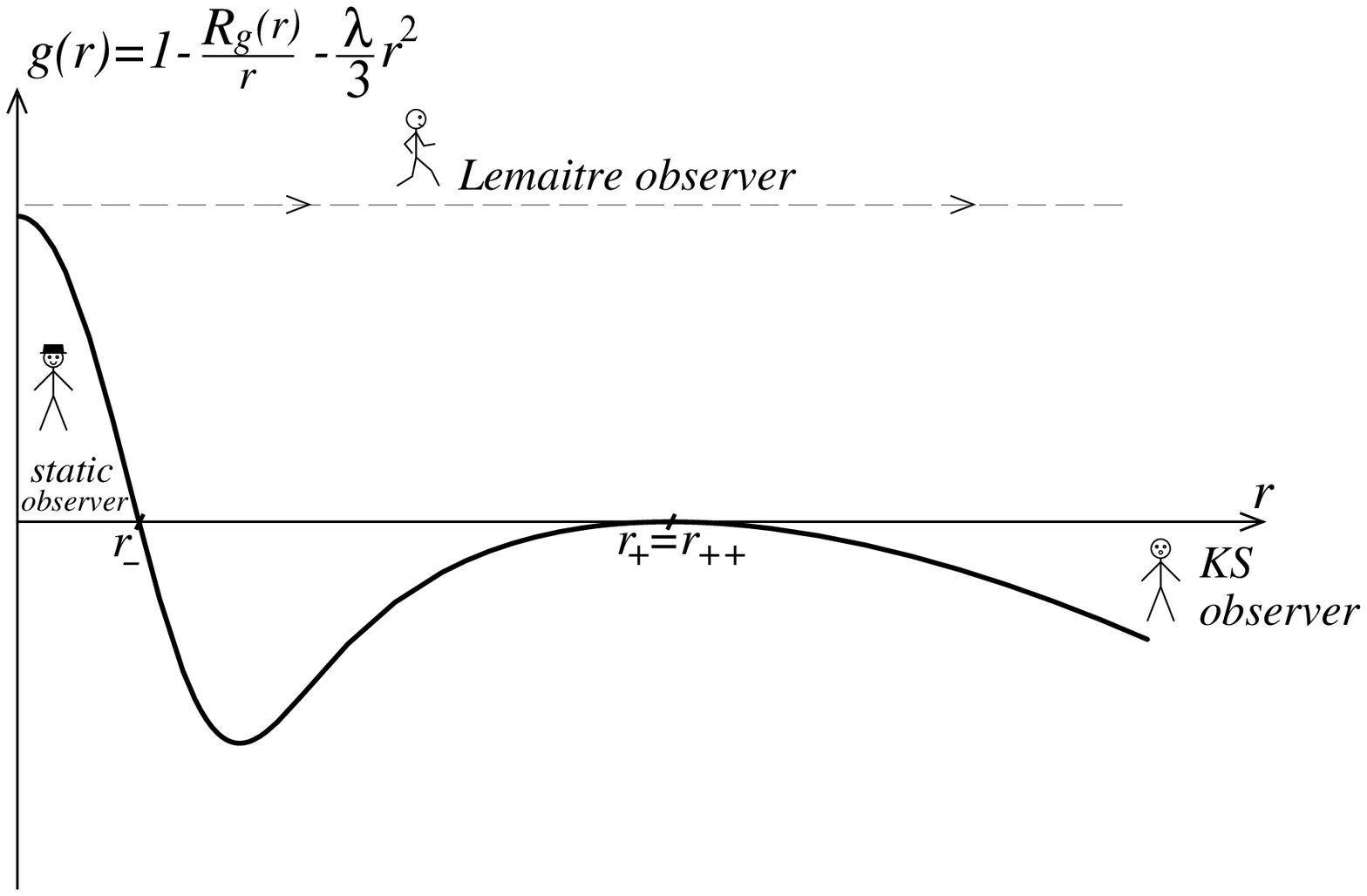,width=8.0cm,height=5.5cm}
\end{center}
\caption{$\Lambda^{\mu}_{\nu}$ geometry with a double horizon
$r_{+}=r_{++}$.}
\label{fig.8}
\end{figure}
Global structure of space-time is shown in Fig.9 \cite{us03}.
\begin{figure}
\vspace{-8.0mm}
\begin{center}
\epsfig{file=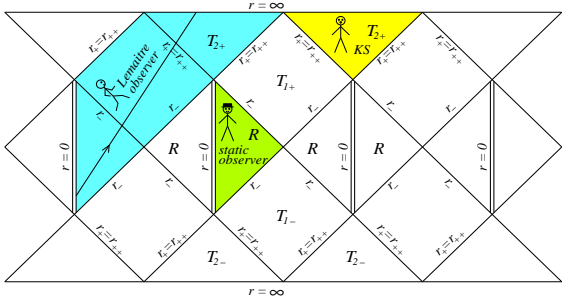,width=8.0cm,height=4.5cm}
\end{center}
\caption{ The global structure of $\Lambda^{\mu}_{\nu}$ space-time
with a double horizon $r_{+}=r_{++}$. }
\label{fig.9}
\end{figure}
 For observers in R-region
 between $r=0$ and $r=r_{-}$, horizon is dominated by the big value
$\Lambda$. The Gibbons-Hawking temperature $T \sim {\hbar c
\sqrt{\Lambda/3}}$.

Coordinate frame of Lemaitre observers map the segment which
contains  R-region and two T-regions.

The surface $r_{+}=r_{++}$ is the null bang  for Kantowski-Sachs
observers in $T_{+}$-regions between $r_{+}=r_{++}$ and infinity,
the bang occurs in their infinitely remote past. Nevertheless,
pre-bang information arrives to Kantowski-Sachs observers, since
for any other geodesics with $E^2 > 1$ in Kantowski-Sachs part of
a manifold, a time for arriving at any finite value of $r$ is
finite \cite{us03}.

 For in-falling KS observers in $T_{-}$ regions, an
apparent coordinate singularity is in their infinitely remote
future.

\vskip0.1in

{\bf Double horizon $r_{-}=r_{+}$}.

\vskip0.1in

This configuration is shown in Fig.10, global structure of
space-time in Fig.11. For a static observer in R-region between
$r_{-}=r_{+}$ this is nonsingular ($\Lambda$ as $r\rightarrow 0$),
cosmological ($\lambda$ as $r\rightarrow \infty$), extreme (double
horizon) black hole.
\begin{figure}
\vspace{-8.0mm}
\begin{center}
\epsfig{file=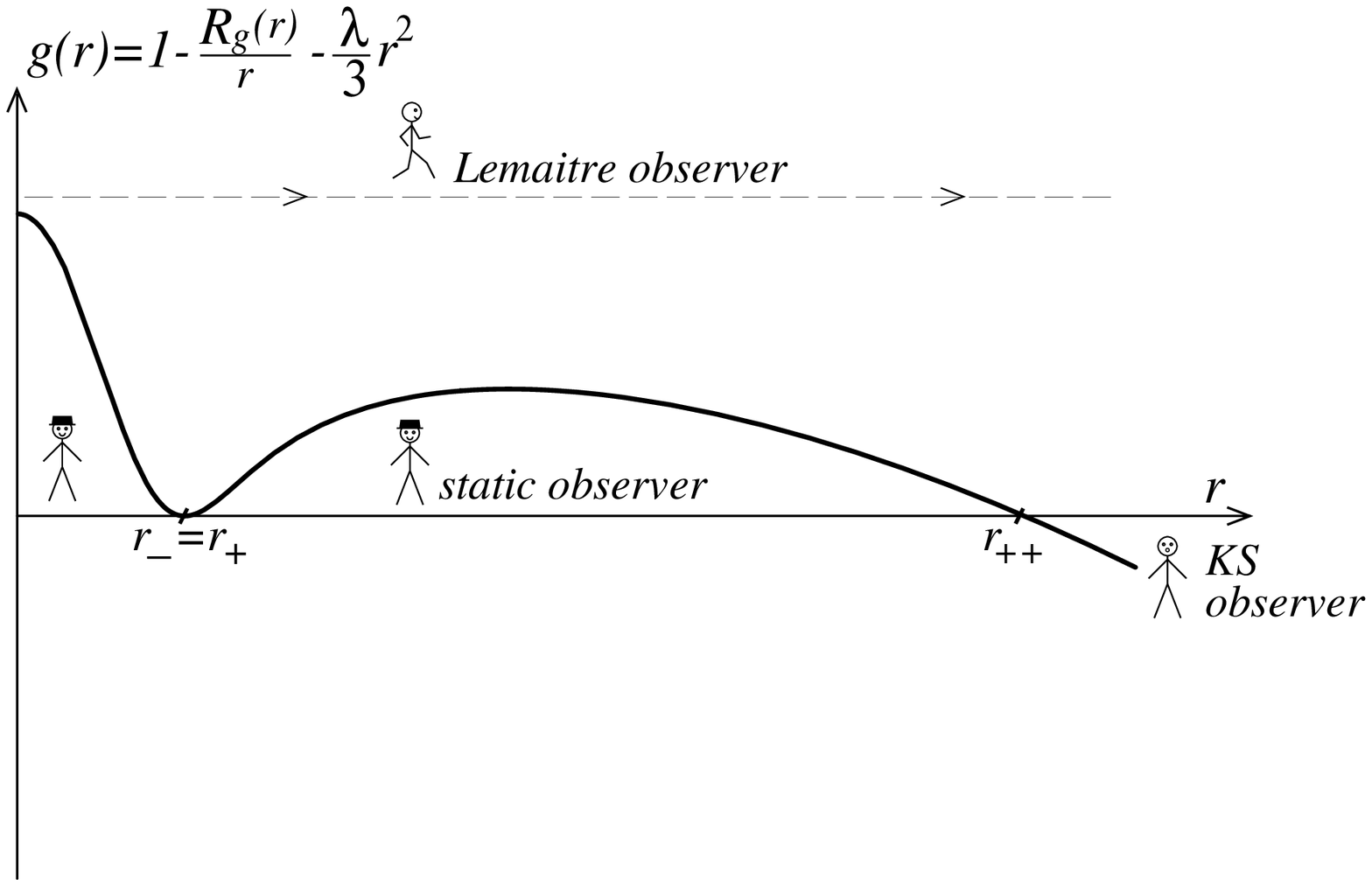,width=7.0cm,height=4.5cm}
\end{center}
\caption{ $\Lambda^{\mu}_{\nu}$ geometry with a double horizon
$r_{-}=r_{+}$. }
\label{fig.10}
\end{figure}
%

\begin{figure}
\vspace{-8.0mm}
\begin{center}
\epsfig{file=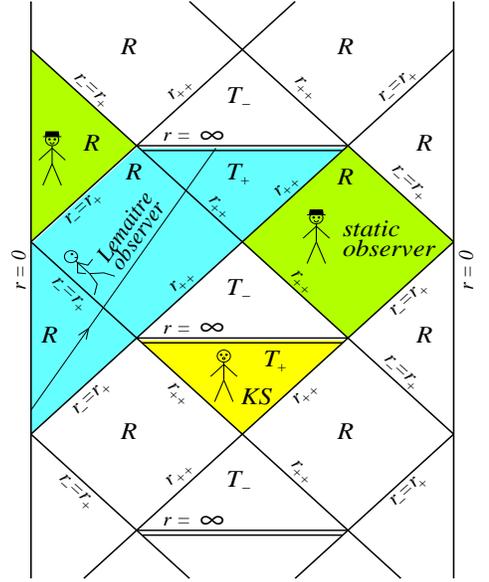,width=7.0cm,height=8.5cm}
\end{center}
\caption{ The global structure  for $\Lambda^{\mu}_{\nu}$
space-time with a double horizon $r_{-}=r_{+}$. }
\label{fig.11}
\end{figure}
This  extreme black hole state has no analogy in singular models,
it appears because of replacing a black hole singularity with an
asymptotically de Sitter core which produces an additional
internal horizon, and a lower bound on a black hole mass comes as
a result \cite{me96}. When a black hole approaches this limit,
Hawking temperature from the BH horizon goes to zero \cite{me96}.
An observer in R-region between $r_{-}=r_{+}$ and $r_{++}$ sees
only Hawking radiation from his cosmological horizon $r_{++}$.

Segment of a manifold available for Lemaitre observers includes
two R-regions and one T-region.

For Kantowski-Sachs observers in T-region evolution starts with a
null bang in finite time in their past.

\section{ Summary and Discussion}

Dependently on the number of horizons, $\Lambda^{\mu}_{\nu}$
geometries describe five types of globally regular configurations,
seen by a static observer as cosmological vacuum nonsingular black
hole with an additional internal horizon separating an additional
R-region, its two extreme states, and two types of one horizon
configurations with global structure of de Sitter geometry but
with vacuum energy $\Lambda^t_t=8\pi G \rho_{vac}$ smoothly
evolving from $\Lambda$ to $\lambda$.

All $\Lambda^{\mu}_{\nu}$ cosmological models belong to the
Lemaitre class of
 models with anisotropic perfect fluid. Cosmological evolution for
 Lemaitre co-moving observers starts with a
non-singular non-simultaneous de Sitter bang which is followed by
a Kasner-type stage of anisotropic expansion. At late times all
models approach de Sitter asymptotic with small $\lambda$.

Spherically symmetric $\Lambda^{\mu}_{\nu}$
 models of Kantowski-Sachs type  contain the regular R-regions.
 For contracting $T_{-}$-regions evolution finishes and for expanding
$T_{+}$-regions starts with the null bang from horizons, which is
in a finite proper time of Kantowski-Sachs observers in the case
of a simple horizon, and in their infinitely remote past in the
case of a higher-order horizon.

Kantowski-Sachs observers get information about pre-bang history
brought by particles and photons which have crossed the horizon in
their finite proper time.

\subsection{Acknowledgment}

This work was supported by the Polish Committee for Scientific
Research through grant No. 5P03D.007.20.

\end{document}